\documentclass[11pt]{article}
\usepackage{graphicx}
\begin{document}

\begin{center}

{\bf \Large Massive circumstellar envelope around 
type IIn supernova SN~1995G}

\bigskip

N.N. Chugai$^{1}$, I.J. Danziger$^{2}$

\medskip

$^{1}$Institute of astronomy RAS, Pyatnitskaya 48, 109017 Moscow, Russia\\
$^{2}$Osservatorio Astronomico di Trieste, via G.B. Tiepolo 11,\\ 
34131 Trieste, Italy\\
\end{center}

\bigskip
\parbox[c]{11.5cm}{\small
We model the interaction of the supernova SN~1995G with a
dense circumstellar (CS) gas in a thin shell approximation.
A model fit of the observed bolometric light curve combined 
with data on the supernova expansion velocity provides 
an estimate of the density of the CS shell, its mass 
($\approx 1~M_{\odot}$), and age ($\approx 8$ years).
It is shown that the derived CS gas density does not depend on the
assumed mass of the supernova envelope. This results from
the high CS density, which ensures that the forward shock 
wave is essentially radiative. The derived CS density is 
consistent with the H$\alpha$ luminosity and with the 
presence of the apparent effect of Thomson scattering in the red 
wing of this line. The mass of the CS envelope together with 
its expansion velocity indicates that the CS envelope was 
ejected as a result of violent energy release 
($\sim 6\times10^{48}$ erg) eight years 
before the supernova outburst.
}

\bigskip

\section{Introduction}

Type IIn supernovae (SN~IIn) with narrow H$\alpha$ emission, 
introduced as a separate family by Schlegel (1990), explode 
in a very dense CS environment. This is demonstrated by the 
presence of a strong CS H$\alpha$, high bolometric 
luminosity and strong broad H$\alpha$ emission powered by the 
CS interaction (Chugai 1990, 1992).  
An analysis of the optical effects of the CS interaction 
provides an efficient diagnostic tool for the CS density 
around SN~IIn. The use of this probe led to the detection 
of the unusually dense CS environment around SN~1987F, 
(Chugai 1990), SN~1997ab (Salamanca et al. 1998), 
SN~1997cy (Turatto et al. 2000) and some other SN~IIn.
For SN~1997ab with the velocity of the CS gas of 
$\approx 90$ km s$^{-1}$ the derived mass loss is enormous,
$\sim10^{-2}~M_{\odot}$~yr$^{-1}$ (Salamanca et al. 1998).
The mechanism for such a tremendous mass loss rate is unclear; 
it exceeds the mass loss rate of the most extreme cases 
of a red supergiant superwind by at least a factor of ten.

The problem of the mechanism of a powerful mass loss rate 
by SN~IIn presupernovae is becoming urgent in view of 
the recent results of the SN~1994W study, which show that 
the CS shell in this case was created by 
mass loss with an average rate of $\sim 0.2~M_{\odot}$ yr$^{-1}$
and the enormous kinetic luminosity, two orders of magnitude 
greater than the radiative luminosity of a massive presupernova
(Chugai et al. 2003). It was suggested there that such a powerful
mass loss by presupernova was related to an explosive 
event about 1.5 yrs before the SN~1994W outburst.
A specific feature of this supenova is the relatively high 
velocity of the CS gas ($u\approx 10^3$ km s$^{-1}$) that 
eventually leads to the energy problem for the superwind mechanism.

Originally, the idea of explosive mass loss several years 
before the supernova explosion was proposed by 
Weaver and Woosley (1979) in connection with a possible 
 strong Ne flash in the degenerate O/Ne/Mg core.
Grasberg and Nadyozhin (1986) suggested explosive ejection of
the presupernova envelope roughly 50 days prior to supernova 
explosion in order to account for the narrow lines in SN~1983K.
However at present the explosive mass ejection by presupernovae 
is just a  working hypothesis, especially keeping in mind 
that Woosley et al (2002) recently expressed doubts about the
reality of their mechanism.
In this respect, the study of signatures of the explosive 
mass ejection by presupernovae (large mass and energy of the CS shell
and small age) can provide us with interesting information 
about poorly undersood phenomena occuring in presupernovae 
on the eve of a supernova explosion.

It was already noted that explosive mass ejection by 
presupernovae might occur in those SN~IIn which show 
high CS velocity ($\sim 1000$ km s$^{-1}$) and  
CS subordinate hydrogen and metal lines (Chugai et al. 2003).
Apart from SN~1994W this type of SN~IIn includes another 
well observed supernova SN~1995G (Pastorello et al. 2002).
This supernova has a specific light curve interpreted as 
a result of CS interaction (Pastorello et al. 2002).

In the present paper we use the model of the 
bolometric light curve in order to extract information about the 
CS gas density. This will permit us to derive mass, energy, and age 
of the CS envelope and thus a conclusion concerning 
its origin. In section 2 we give a brief description of the model, 
in section 3 we explore model sensitivity to parameters 
and demonstrate the uncertainty of the parameters recovered in the case 
of SN~1997cy. In section 4 we model the light curve of SN~1995G 
and derive the mass of the CS envelope. The discussion of results
and their relationship to the H$\alpha$ intensity and profile 
is presented in section 5.

The paper is based upon the photometry and spectra 
presented by Pastorello et al. (2002). We use the 
Hubble constant of $70$~km~s$^{-1}$~Mpc$^{-1}$.

\section{Model}

The light curve model applied below was described earlier 
(Chugai 2001). Here we only briefly recapitulate its essential 
features. We consider the expansion of the 
supernova in the CS enevelope with a given density 
distribution and velocity in the thin shell approximation 
(Chevalier 1982a). We neglect the structure of the 
interaction region which consists of the two shock waves,
forward and reverse, with a density peak in between. This 
structure may be described by self-similar solution under 
proper conditions (Nadyozhin 1981, 1985; Chevalier 1982b).
We are interested in the more general case of density distributions 
with non-zero CS velocity when the self-similar solution is not 
applicable, so the dynamics of the thin shell will be calculated
 numerically.

\begin{table*}
\caption{Model parameters}

\bigskip
\begin{tabular}{cccccccc}
\hline

 & $M$         & $E$           & $s$ & $w_0$ & $\rho_0$            
     & $R_{\rm b}$ & $M_{\rm cs}$ \\
       & $M_{\odot}$ & $10^{51}$ Œerg &  &  $10^{17}$ g cm $^{-1}$  & 
       $10^{-15}$ g cm$^{-3}$ &  $10^{16}$ cm &  $M_{\odot}$ \\
\hline

  A1  &  5   &   1    &   1    & 0.4   & 3.2  & 1    &  1  \\
  A2  &  5   &   0.1  &   1    & 0.4   & 3.2  & 1    &  1  \\
  B   &  1   &   1    &   1    & 0.4   & 3.2  & 1    &  1  \\
  C   &  5   &   1    &   2    &   2   & 16   & 1    &  1  \\
  D   &  5   &   1    &   1    & 0.8   & 6.4  & 1    &  2  \\
  cy1 &  5   &   2    &   1.8  & 1     & 8    & 2.3  & 1.8 \\
  cy2 &  1.5 &   1    &   1.3  & 1.1   & 9    & 1.7  & 4.1 \\
  G1  &  2   &  0.24  &   2    & 1.2   & 9.5  & 2.2  & 1.2 \\
  G2  &  10  &  0.6   &   2    & 1.1   & 8.7   & 2    &  1.1 \\
  G3  & 21   &   1    &   2    & 1.2   & 9.5  & 2    & 1.1 \\  
\hline
\end{tabular}
\end{table*}

The interaction of a supernova with its CS environment on 
a time scale greater than several days weakly depends on the 
initial epoch of the interaction. We assume the following: 
the interaction begins at 
the presupernova radius $R_0$ and the supernova expands freely 
($v=r/t$), and its density distribution is an inner plateau with the outer 
power law density drop ($\rho \propto v^{-9}$).
The numerical solution of the equation of motion with the final 
velocity of the CS gas provides the thin shell radius 
$R(t)$, and the relative velocities of the supernova and CS gas 
flows. These permit us to calculate the kinetic luminosity of the 
forward and reverse shock, which may be transformed into 
X-ray luminosities of both shocks and eventually into 
optical bolometric luminosity (Chugai 1992).
The contribution of the luminosity supplied by the 
internal energy stored in the supernova during the explosion is 
calculated according to an analytical approximation (Arnett 1980, 1982).
In our model the full light curve is a linear 
superposition of this luminosity and the interaction luminosity.
This approach permits us to take into acount the radiation 
of the initial internal energy at the early epoch 
in a straighforward way.

\begin{figure}[t]
\centering\includegraphics[scale=0.6,angle=0]{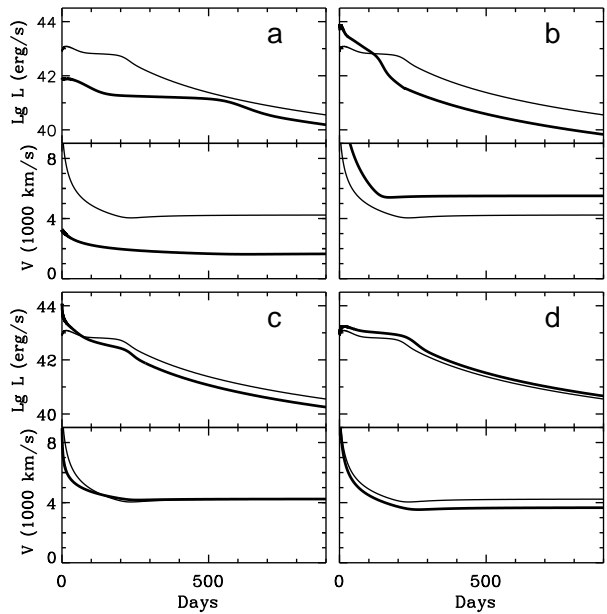}
\caption[]{\footnotesize Bolometric light curves and velocities of the 
thin shell for models A--D (Table 1). 
Model A1 (thin line) is a template for other models 
A2, B, C, D, which are shown in panels a, b, c and d, respectively.
 }
\label{f:dem}
\end{figure} 

The CS density is described by the density $\rho_0$ at the 
radius $10^{15}$ cm, or density parameter $w=4\pi r^2\rho$ at 
this radius, and by the power index ($s$) in the power law 
($\rho\propto r^{-s}$).
The extent of the CS envelope is characterized by the outer 
radius $R_{\rm b}$. With the mass ($M$) and energy ($E$) 
of the supernova envelope we have five parameters, which are 
constrained by the light curve, photospheric radius, 
supernova expansion velocity and phase of the late rapid light 
curve decline.

\section{Parameter sensitivity and SN~1997cy}

The sensitivity of the model to parameter variations is demonstrated by 
the models A1, A2, B, C, D (Fig. 1) with parameters 
shown in Table 1. The table presents beginning from 
the second column: supernova mass ($M$), energy ($E$), power 
index of the CS density distribution ($s$), density parameter 
($w_0$) and density ($\rho_0$) at the radius $10^{15}$ cm, 
outer radius of the CS envelope ($R_{\rm b}$), and the mass of 
the CS envelope ($M_{\rm cs}$). The latter is the output value
and not a free parameter.
We adopt the CS velocity 1000 km s$^{-1}$,
presupernova radius $R_0=1000~R_{\odot}$, and $^{56}$Ni mass 
$0.003~M_{\odot}$ following the estimate for SN~1994W 
(Sollerman et al. 1998). The model A1 is adopted as a 
standard, with which all the other models are compared.
For all the cases Fig. 1 shows the bolometric 
light curves and the thin shell velocities.

\begin{figure}[t]
\centering\includegraphics[scale=0.5,angle=0]{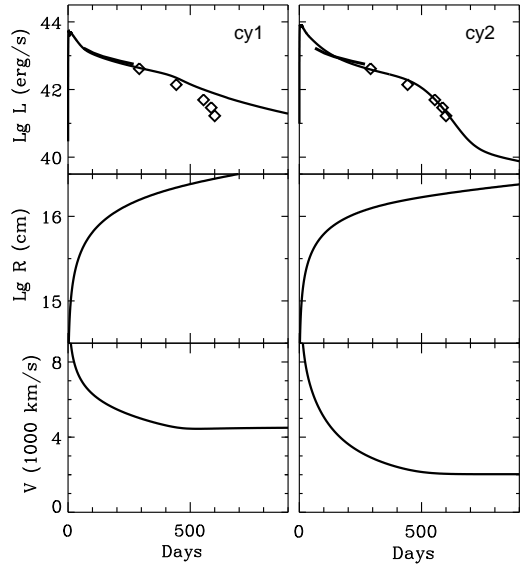}
\caption[]{\footnotesize Two models of bolometric light curve 
of SN~1997cy (Table 1). Observations (diamonds and thick line cut)
are taken from Turatto et al. (2000). Modeling shows that 
the light curve may be reproduced for the moderate value of 
supernova energy. 
 }
\label{f:cy}
\end{figure}

We begin with a brief discussion of the general properties of models.
First, they all show a shoulder, which reflects the   
overtaking of the CS shell boundary by the forward shock. 
Second, after that time the luminosity is still high.
The source of this radiation is the inner shock which 
is driven by the supersonic velocity jump between the outer
supernova material and the thin shell. As the
thin shell is accelerated, the inner shock luminosity 
decreases.
Note that the luminosity of the inner shock is lower 
in the case of a lower mass of the supernova envelope 
(model B).
Finally, for most models, the contribution of the 
internal energy of the supernova to the luminosity is 
relatively small, except for the model A2 with a
noticeable early bump during the first $50-100$ days.

Now let us consider the effects of parameter variations.
The reduction of the supernova kinetic energy by an 
order of magnitude reduces the early luminosity by more than 
an order of magnitude with a less pronounced luminosity decrease 
at the late epoch (Fig. 1a). The five-fold mass reduction 
(model B) results in the substantially higher early luminosity 
due to the higher velocity of the outer shock (Fig. 1b). Still the 
 model B shows a faster decay of the early luminosity because of 
the fast crossing of the CS envelope. 
The model C with the steeper density distribution ($s=2$) for a
similar mass of the CS shell gives, as expected, a higher 
early luminosity and faster decay (Fig. 1c).
The model D with twice as higher mass of the CS envelope 
results in the higher luminosity and lower velocity of the 
thin shell (Fig. 1d).

\begin{figure}[t]
\centering\includegraphics[scale=0.6,angle=0]{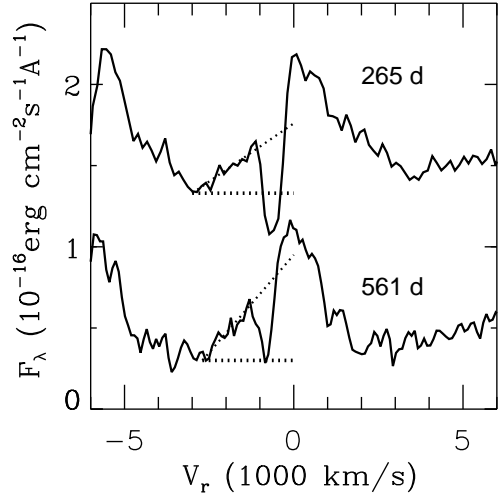}
\caption[]{\footnotesize Fe II 5018 \AA\  line in SN~1995G spectra on
days 265 and 561 (Pastorello et al. 2002). Both profiles 
reveal narrow P Cyg component and broad emission component. 
The latter is related to the dense shell at the supernova boundary.
Dotted line shows approximation of the blue part of 
profile and continuum used for the estimate of the maximal velocity.
 }
\label{f:sp}
\end{figure}

The above analysis implies that a combination of low mass and 
low energy is able to produce the rather bright SN~IIn phenomenon 
as a result of the efficient deceleration of the bulk of the 
supernova envelope. 
As an ilustration of this statement let us consider the 
light curve of SN~1997cy (type IIn), studied by 
Turatto et al. (2000). In the cited paper the proposed interaction model 
suggests the large supernova energy ($3\times10^{52}$ erg) and 
mass ($\sim20~M_{\odot}$). Although this model is plausible,
one cannot rule out an alternative model with 
moderate values of both energy and mass. This is demonstrated by models 
cy1 and cy2 (Fig. 2 and Table 1). The adopted velocity of the CS envelope 
is 10 km s$^{-1}$. Model cy1 with 
$M=5~M_{\odot}$ and $E=2\times10^{51}$ erg provides a satisfactory 
agreement with data, while the model cy2 with 
$M=1.5~M_{\odot}$ and $E=10^{51}$ erg shows an even slighly better 
fit since it better reproduces the luminosity drop at $t>600$ days.
The velocity of the thin shell is different in these models and 
such a difference may be crucial in discarding inappropriate 
models on the 
basis of the velocity information provided by the line profiles. 
Unfortunately, there is no straighforward procedure for determining 
 the thin shell velocity in SN~1997cy. So the uncertainty of the 
choice of the mass and energy remains.

\section{SN~1995G light curve and the mass of CS envelope}

Let us briefly consider additional observational constraints 
on the model apart from the bolometric light curve of SN~1995G.
The energy distribution in early spectra of SN~1995G provides
the estimates of temperature and radius of the photosphere
on days 2 and 36 (Pastorello et al. 2002).
We believe that the photospheric radius at the early epoch is 
approximately equal to the radius of the thin shell.
The arguments are based upon the result that the thin dense shell 
is opaque in the optical band in the case of a very dense 
CS environment ($w\sim10^{17}$ g cm$^{-1}$) for about 
one-two months (Chugai 2001). 
Thus, the first additional observational constraint on the model 
suggests that the early photospheric radius should approximately 
coincide with the radius of the thin shell, if the CS density 
is rather high, i.e. $w\sim10^{17}$ g cm$^{-1}$.

The next constraint of the model is provided by the expansion 
velocity of the thin shell. The observational information about 
this velocity should be extracted from the line profiles, 
particularly from the maximal velocity of the broad component.
Since at the early epoch Thomson scattering can 
contribute in the broad wings (Chugai et al. 2003), in order
 to estimate the thin shell velocity we rely on 
the late nebular spectra of SN~1995G on days 265 and 561. 
In Fig. 3 the Fe II 5018 \AA\ line is shown for both epochs 
(Pastorello et al. 2002). This line is free from blending 
which makes it a reliable indicator of the broad component 
velocity. The broad component is identified with the 
dense thin shell, possibly partially fragmented, at the 
boundary between the supernova and the CS gas as in SN~1994W 
(Chugai et al. 2003). The maximal velocity in the 
blue wing of the broad component is estimated using a simple 
procedure of the linear approximation of the line and 
continuum flux (Fig. 3). The derived maximal velocities are 
3000 km s$^{-1}$ and 2700 km s$^{-1}$ on days 265 and 561, 
respectively, with a possible uncertainty of 10\%. 
We attribute these velocities to the thin shell. 
Similar values ($\sim 3000$ km s$^{-1}$) are shown by the
 maximal velocity 
in the H$\alpha$ blue wing. However, in this case Fe II emission 
lines may contribute to the blue wing flux, which may hamper 
the reliability of the estimate of the maximal velocity.

The expansion velocity of the CS envelope estimated from narrow
absorption lines is $u=750$ km s$^{-1}$ according to 
Pastorello et al. (2002). However, a somewhat larger value 
$u\approx 850$ km s$^{-1}$ is demonstrated by the emission lines 
of the infrared triplet of Ca II (Pastorello et al. 2002, their Fig. 10).
We adopt here $u=800$ km s$^{-1}$. 

\begin{figure}[t]
\centering\includegraphics[scale=0.5,angle=0]{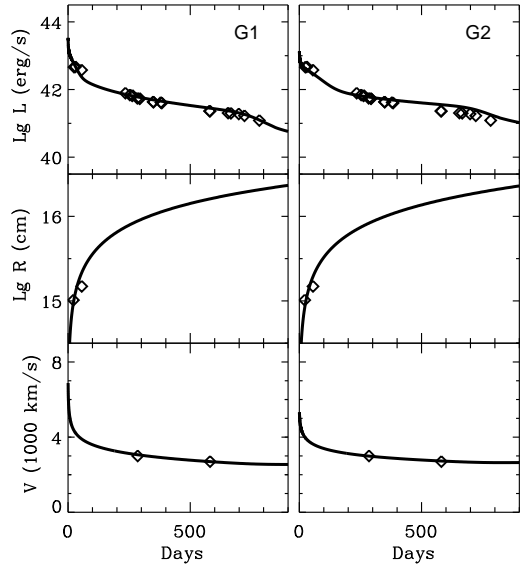}
\caption[]{\footnotesize
Bolometric light curve of SN~1995G (upper panel) 
radius of the thin shell (middle) and the velocity of the thin shell 
(lower panel).  Shown are two models  (lines) G1 (left) and G2 (right) 
(see parameters in Table 1). Diamonds are empirical values taken from 
Pastorello et al. (2002) except for velocities, which are measured 
in the present paper. 
 }
\label{f:g12}
\end{figure} 

Finally, yet another constraint on the model is the observation 
that after day 700 the light curve shows a more rapid decay which 
is interpreted as the result of the overtaking of the outer boundary of the 
CS envelope by the forward shock (Pastorello et al. 2002).
This fact will  be used to estimate the outer radius of 
the CS envelope.

Preliminary computations of an extended set of 
the expansion dynamics and bolometric light curves for SN~1995G
reveal the following important feature: it emerges that 
within empirical constraints the model is 
not sensitive to either of the two guiding supernova parameters, 
mass and energy.
Selecting mass as a primary parameter, we found that 
a fit is achievable for a wide range of mass values.
Two models with masses $2~M_{\odot}$ (model G1) and 
$10~M_{\odot}$ (model G2), with other parameters given in Table 1, show an
acceptable fit of the bolometric light curve, photospheric radius and 
thin shell velocity (Fig. 4). Note that to reach agreement with 
 the photospheric radius in the first epoch we added 
 20 days to the age of the supernova given by Pastorello et al. (2002)
 which means that the explosion of SN~1995G is assumed to occur 
 20 days earlier than the zero point accepted in the cited paper.
However, when some observational phase is mentioned in the 
text, we retain formally its day according to Pastorello et al. (2002).
The models G1 and G2 confirm that the density and mass of the 
CS envelope do not depend on the adopted supernova 
mass (Table 1). The density at the radius of $10^{15}$ cm is 
$\approx 9\times10^{-15}$ g cm$^{-3}$, or in terms of the hydrogen 
concentration for a normal abundance, $n\approx 4\times10^9$ cm$^{-3}$.

The independence of the CS gas density on the adopted supernova 
mass has a simple explanation. For the high CS density required 
in case of SN~1995G both shock waves (forward and reverse) are 
essentially radiative during a long period ($\approx700$ d) which 
means that for a given total radiated energy the amount of the 
dissipated kinetic energy in shocks is invariant.
Since the forward shock dominates in the luminosity, the 
latter suggests that the overall radiated energy should be 
of the order of the average value of $0.5M_{\rm cs}(v-u)^2$, which 
thus must be invariant.
Given observational constraints on the shell velocity ($v$) and 
CS velocity ($u$), the total mass of the CS envelope, thus, 
must also be invariant in different models.

In both models the kinetic energy is lower than the typical energy 
of core collapse supernovae ($10^{51}$ erg). Although the 
question of the typical value of the energy for SN~IIn is open,
it would be instructive to consider a case of a "standard" 
energy.  The model G3 with the energy 
 $E=10^{51}$ erg shows a sensible fit and again requires the 
 similar CS shell density and mass (Table 1) thus supporting 
 the independence of these values on the adopted 
 supernova mass at least in the range of $2-20~M_{\odot}$. 
Note, the latter statement can be reformulated in terms of 
supernova energy as a guiding parameter. In that case the derived
CS density around SN~1995G is independent of the supernova energy 
at least in the range  of $(0.24-1)\times10^{51}$ erg.

\begin{figure}[t]
\centering\includegraphics[scale=0.5,angle=0]{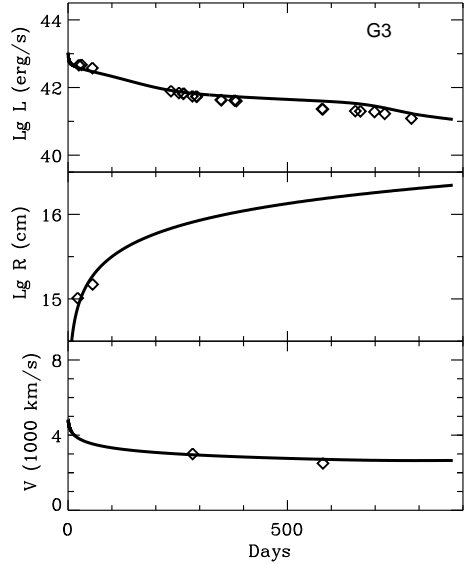}
\caption[]{\footnotesize The same as in Figure 4 but for the model G3.
 }
\label{f:g3}
\end{figure}

The fact of the weak dependence of the CS gas density on 
the adopted supernova mass in the interaction model
is of importance for the diagnostics of the CS density 
around SN~IIn. The model of the bolometric light curve 
in combination with the velocity of the thin shell 
thus essentially provides us with a confident estimate of the CS 
density in the limit of the radiative forward shock wave.
This tool was already used earlier for SN~1987F 
in an assumption of the standard supernova energy (Chugai 1992). 
Now it has become clear that in the limit of high 
CS density ($w\geq 10^{17}$ g cm$^{-1}$) the derived 
CS density practically 
does not depend on the adopted supernova mass, at least 
in the range of $2\leq M\leq 20~M_{\odot}$.

\section{Discussion}

The CS gas density distribution found above has direct 
implications for the interpretation of the spectrum 
of SN~1995G.
The similarity of early spectra of SN~1995G and SN~1994W,
particularly the presence of the strong effect of Thomson 
scattering in the H$\alpha$ profile (Chugai et al. 2003) 
suggests that a substantial fraction of H$\alpha$
is emitted by the CS enevlope. Is this picture consistent 
with the above CS density estimate?

The H$\alpha$ luminosity of the CS enevelope in the range 
of $r_1<r<r_2$ in our model ($\rho\propto r^{-2}$) is 

\begin{equation}
L({\rm H}\alpha)= \frac{1}{4\pi r_1} \alpha_{32}h\nu_{23}(xXwN_{\rm A})^2
\left(1-\frac{r_1}{r_2}\right)\,,
\label{rha}
\end{equation}

\noindent wher $\alpha_{32}$ is the effective recombination 
coefficient of H$\alpha$ emissivity, $h\nu_{23}$ is the energy of the
H$\alpha$ photon, $x$ is the ionization degree, $X$ is the 
hydrogen abundance, $N_{\rm A}$ is the Avogadro number. Substituting 
in equation (\ref{rha}) the values for day 2, 
namely, the inner radius 
$r_1$ equal to the radius of the photosphere $1.1\times10^{15}$ cm
(Pastorello et al. 2002),
outer radius $r_2=R_{\rm b}=2\times10^{16}$ cm, 
$\alpha_{32}=1.2\times10^{-13}$ cm$^3$ s$^{-1}$ (for 
electron temperature $10^4$~K), $w=1.2\times10^{17}$ g~cm$^{-1}$,
one obtains $L({\rm H}\alpha)=6.2\times10^{40}x^2$ erg s$^{-1}$ 
assuming $X=0.7$.
On the other hand, the observed H$\alpha$ luminosity on day 2 is 
$L({\rm H}\alpha)=6.3\times10^{40}$ erg s$^{-1}$ (Pastorello et al. 2002).
Following the SN~1994W case we assume that the contribution of the 
CS component in H$\alpha$ is at least half of the total line luminosity
(Chugai et al. 2003). Comparing the model and observational luminosity 
of the CS component we thus conclude that both values agree,
if the average ionization degree of the model CS envelope 
is $0.7<x\leq1$.

The optical depth of the CS envelope to Thomson scattering 
for the same epoch is 
$\tau_{\rm T}=k_{\rm T}wx/4\pi r_1\approx 2.6x$. For the 
above range of the ionization degree this gives 
$1.8<\tau_{\rm T}<2.6$. The presence of the strong red wing 
in H$\alpha$ on day 2 (Pastorello et al. 2002, their Fig. 10) 
suggests significant Thomson scattering, so one expects $\tau_{\rm T}>1$.
Moreover, close similarity of this profile with the H$\alpha$ in SN~1994W
on day 30 (Chugai et al. 2003, their Fig. 10) suggests that
the optical depth of the CS envelope in SN~1995G is conceivably 
as large as $\tau_{\rm T}\sim 2$, which is consistent 
with the above range of Thomson optical depth.

To summarize, the density of the CS envelope derived from the 
interaction model agrees with both the H$\alpha$ line luminosity and 
the presence of strong effects of Thomson scattering in this line.
 
Generally, analysis of CS absorption lines might provide us 
additional information about the CS density. However this approach
would require a rather complicated model of ionization and excitation
in the CS envelope.
Simple considerations based, for instance, on Fe II absorption, 
provide a rough lower limit (Sollerman et al. 1998). 
The requirement that absorption line with the optical depth 
$\tau$ is present in the spectrum is $\tau>1$. For an envelope 
of a size $r$ with the average velocity dispersion on the line of sight 
of the order of the expansion velocity $v$ (greater than 
thermal velocity), the latter condition is  

\begin{equation}
\tau\approx\sigma(\nu)n_1r\approx\sigma_0n_1r\lambda_{12}/v >1\,,
\label{fe}
\end{equation}

\noindent where $\sigma_0=(\pi e^2/m_ec)f_{12}$ is
 the integrated over frequency cross-section $\sigma(\nu)$, 
 $f_{12}$ is the oscillator strength, $n_1$ is the concentration 
 at the lower level, $\lambda_{12}$ is the wavelength. 
For the Fe II 5018 \AA\ absorption ($f_{12}=0.01$) 
adopting on day 2 the excitation 
temperature of the lower level equal to the photospheric 
one, 8800 K (Pastorello et al. 2002), and taking $r$ equal 
to the photospheric radius ($1.1\times10^{15}$ cm) one gets 
the lower limit of the hydrogen concentration assuming a solar 
Fe abundance $n>3\times10^7$ cm$^{-3}$ in qualitative agreement 
with the density found from the light curve analysis.

The outer boundary of the CS enevelope ($2\times10^{16}$ cm) combined 
with the CS expansion velocity 800 km s$^{-1}$ implies the 
age of the CS envelope $t_{\rm cs}\approx 8$ yr which is close to the 
estimate of the starting time for the strong mass loss, 
$\approx12$ yr before the supernova explosion found by 
Pastorello et al. (2002). The mass of the CS envelope 
($1~M_{\odot}$) combined with the age thus suggests the average 
mass loss rate $\dot{M}\sim 0.1~M_{\odot}$~yr$^{-1}$ which is an enormous
 value. The estimated total kinetic energy of the CS enevelope 
is $E_{\rm cs}\approx 6\times10^{48}$ erg and the average kinetic 
luminosity of the mass loss is then 
$E_{\rm cs}/t_{\rm cs}\approx2.4\times10^{40}$ erg s$^{-1}$.
This value is almost a factor of two in excess of the typical radiative 
luminosity of a massive presupernova ($\approx10^5~L_{\odot}$).
Thus the mass loss certainly cannot be attributed to the superwind.

We propose, therefore, that the mass ejection in the SN~1995G presupernova 
was initiated by some powerful energy release in the hydrodynamic time
scale approximately 8 years before the major supernova explosion.
In fact we are reproducing here arguments used in the case of SN~1994W
to conclude that the CS enevelope around SN~1994W was lost as a 
result of an explosive 
event $\sim 1.5$ yr before the supernova explosion (Chugai et al. 2003).
It was proposed there that the 
explosive mass ejection was initiated by the flash of nuclear 
burning of Ne in the degenerate O/Ne/Mg core. This assumption 
follows the original hypothesis of Weaver and Woosley (1979) concerning the 
behavior of presupernovae with initial masses 
of $\approx 11~M_{\odot}$. A similar possibility might occur 
also in the case of SN~1995G. Note that the age of the CS envelope 
around SN~1995G ($\approx 8$ yr) lies within the range of 
the phase for the Ne burning in massive star cores $1-10$ years 
before the supernova explosion (Heger 1998).

If the initial mass of the SN~1995G presupernova was actually close 
to $11~M_{\odot}$ then given a neutron star of $1.4~M_{\odot}$ 
the supernova ejecta cannot exceed $10~M_{\odot}$. 
In that case our interaction model predicts that the 
supernova  kinetic energy $\leq 6\times10^{50}$ erg 
(Table 1). This is lower than the value $1.5\times10^{51}$ erg 
adopted for SN~1994W (Chugai et al. 2003). It may well be that 
the differences of supernova energies and ages of CS shells 
of SN~1995G and SN~1994W are possibly related 
to slight differences in their initial masses or in slightly 
different evolutionary histories.

If the mass ejection of presupernova SN~1995G had an explosive 
nature then the CS envelope expansion regime must be close 
to a free expansion ($u\propto r$) law, at least in the outer layers.
In this respect the increase of velocity derived from
 CS absorption lines of Fe II between days 330 and 560 
 (Pastorello et al. 2002) is qualitatively consistent with the 
possible free expansion CS kinematics.

The envelope ejection with the mass of $\approx1~M_{\odot}$ and 
kinetic energy of $\sim 6\times10^{48}$ erg should be 
accompanied by the optical flash eight years before the explosion
of SN~1995G.
In a simple analytical model of the light curve (Chugai 1991) 
assuming presupernova radius $100<R_0<1000~R_{\odot}$ we
estimate the absolute magnitude of flash maximum as 
 $-12.5 > M_V > -13.5$ mag. with the duration of the light 
 curve of $80-120$ days. For the distance of 63 Mpc the 
 apparent magnitude of maximum for this flash is $21.8>V>20.8$ mag.
The only available image of the host galaxy NGC1643 close 
to the suggested time of the presupernova flash is a UK Schmidt plate 
taken 12 Dec. 1982, 13 years before the SN~1995G explosion.
Therefore this image does not constrain the presupernova flash history.
Inspection of this plate reveals no object brighter than 20.5 mag. 
in J band.

\section{Conclusion}

We performed the modeling of the bolometric light curve and
  expansion dynamics of SN~1995G in the dense CS environment.
As a result we obtained the density and 
mass of the CS envelope which do not depend on the adopted 
mass of the supernova envelope. The derived mass of the CS 
envelope combined with the velocity of the CS gas leads us to 
conclude that the CS envelope was ejected as a result of 
energetic hydrodynamical process eight years before the 
explosion of SN~1995G. We speculate that this mass ejection 
was initiated by the powerful flash of nuclear burning in the 
O/Ne/Mg core of presupernova following the earlier 
conjecture proposed in the case of SN~1994W.

\bigskip
We thank the UK Schmidt Unit for providing scanned images of Schmidt 
plates around NGC1643. The research by N. Chugai is supported by 
RFBR grant 01-02 16295.

\bigskip 
\bigskip
{\bf \large References}

\bigskip

\noindent 1. Arnett W.D., {\em Supernovae:  A survey of current research,} Ed.\\
\indent M.J. Rees, R.J. Stoneham (Reidel, 1982), p. 221.\\
2. Arnett W.D., Astrophys. J. 237, 541 (1980).\\
3. Chevalier R.A., Astrophys. J. 259, 302 (1982a).\\
4. Chevalier R.A., Astrophys. J. 258, 790 (1982b).\\
5. Chugai N.N., Sov. Astr. Lett. 16, 457 (1990).\\
6. Chugai N.N., Sov. Astr. Lett. 17, 210 (1991)\\
7. Chugai N.N., Sov. Astr. 36, 63 (1992)\\
8. Chugai N.N., Mon. Not. R. astron. Soc. 326, 1448 (2001).\\
9. Chugai N.N., Blinnikov S.I., Lundqvist P. {\em et al.},
    Mon. Not. R. astron. \\
\indent    Soc. (in preparation) (2003).\\
10. Falk S.W. and Arnett D.W., Astrophys. J. Suppl. 33, 515 (1977).\\
11. Grasberg E.K., Nadyozhin D.K., Sov. Astr. Lett. 12, 68 (1986).\\
12.  Heger A., {\em The presupernova evolution of rotating massive 
stars}, \\ 
\indent Ph.D dissertation, MPA 1120 (1998).\\
13.  Nadyozhin D.K., Preprint ITEP No. 1 (1981).\\
14.  Nadyozhin D.K., Astrophys. Space. Sci. 112, 225 (1985).\\
15.  Pastorello A., M. Turatto, S. Benetti {\em et al.},
 Mon. Not. R. astron. Soc. \\ 
\indent 333, 27 (2002).\\
16. Salamanca I., Cid-Fernandes R., Tenorio-Tagle G. {\em et al.}, 
 Mon. Not.\\
 \indent R. astron. Soc. 300, L17 (1998).\\
17.  Schlegel E., Mon. Not. R. astron. Soc. 244, 269 (1990).\\
18. Sollerman J., Cumming R.J. and Lundqvist P., 
    Astrophys. J. 493, \\
 \indent   933 (1998).\\
19. Turatto M., Suzuki T., Mazzali P.{\em et al.}, Astrophys. J.
   534, L57 (2000).\\
20. Weaver T.A. and Woosley S.E., BAAS 11, 724 (1979).\\
21. Woosley S.E., Heger  A. and  Weaver T.A., Rev. Mod. Phys. 
74,\\
\indent 1015 (2002).\\

\
\end{document}